\documentclass[runningheads]{llncs}
\usepackage{graphicx}
\usepackage{hyperref}
\usepackage[utf8]{inputenc}
\usepackage[T1]{fontenc}
\usepackage{xcolor}
\usepackage{booktabs}  

\usepackage[font=small,labelfont=bf,tableposition=top]{caption}

\DeclareCaptionLabelFormat{andtable}{#1~#2  \&  \tablename~\thetable}

\usepackage{makecell}
\usepackage{xspace}

\newcommand{\myparagraph}[1]{\paragraph{\textbf{#1}}}

\begin{document}
\title{Alternative Scoring Method of Pleurisy in Slaughtered Pigs: Preliminary Investigations}
%
\titlerunning{Alternative Scoring Method of Pleurisy in Slaughtered Pigs}
\author{
Giuseppe Marruchella\inst{1} \and Michael Odintzov Vaintrub\inst{1} \and Andrea Di Provvido\inst{1} \and Elena Farina\inst{1} \and Giorgio Fragassi \and Giorgio Vignola\inst{1}}

\authorrunning{G. Marruchella et al.}
\institute{University of Teramo, Faculty of Veterinary, Località Piano d’Accio, 64100, Teramo, Italy\\}
\maketitle
\begin{abstract}
The slaughterhouse is very useful in monitoring the health status of livestock, the profitability of their
breeding and the effectiveness of therapeutic and/or prophylactic strategies. Over the years, a number
of methods have been developed to quantify lesions - especially those effecting the respiratory tract,
observed in slaughtered animals. Among these is the “Slaughterhouse Pleurisy Evaluation System”
(SPES), which is widely used to score pleural lesions caused by Actinobacillus pleuropneumoniae. The
aim of the present study is to develop and assess an alternative method to score pleurisy in slaughtered
pigs, based on the inspection of the parietal pleura. This method has been compared with the SPES
grid, which is considered as the “gold standard” in this field of study. Preliminary data indicate that the
two methods provide almost overlapping results, showing very high correlation coefficients. Scoring
pleurisy on the parietal pleura proved to be fast and easy, it could represent a valuable alternative to the
SPES method.

\keywords{pig \and slaughterhouse \and pleurisy \and scoring systems.}
\end{abstract}
\section{Introduction}
\label{sec:introduction}
The slaughterhouse represents an extremely useful observation point for the monitoring the health
status of livestock and providing feedback regarding veterinary and management activities (i.e.
vaccinations, treatments, etc.) performed over the animal’s lifetime. This is particularly relevant in
pigs: the short productive cycle of this species does not permit the full healing of lesions, which are
therefore still detectable at post mortem inspection. Over the years, many systems have been developed
which quantify the impact of diseases such as; sarcoptic mange, atrophic rhinitis, and ascaridiosis.
Despite this, the most importance has always been given to diseases of the lower respiratory tract
(pneumonia caused by Mycoplasma hyopneumoniae, pleuropneumonia caused by Actinobacillus
pleuropneumoniae) which have an important economic impact on pig farming~(\cite{martelli2013patologie,scollo2017benchmarking}). Regardless of the disease considered, the lesion evaluation system must respect
certain criteria a) it must be simple and permit evaluation at the speed of slaughter, b) it should be easy
to standardise and reproduce c) it should produce results which are easily interpretable and can be
analysed statistically \cite{martelli2013patologie}. Pleurisy is a common finding in pigs and is frequently seen at
post mortem inspection of carcasses. Over the years, various scoring systems have been developed for
pleurisy, including the “Slaughterhouse Pleurisy Evaluation System”~\cite{dottori2007proposta}. This
system adequately responds to the above-mentioned criteria and is commonly used to quantify lesions
caused by infections of Actinobacillus pleuropneumoniae~\cite{sibila2014comparison,merialdi2012survey,martelli2013patologie}. This study aims to develop and evaluate an alternative method for scoring pleurisy
in slaughtered pigs, through the inspection of the parietal pleura. This method was compared with the
SPES system which is the most widely used method in field conditions.

\section{Materials and Methods}
\label{sec:method}
\myparagraph{Animals}
A total of 216 pigs, between 9-11 months of age and weighing between 140-180 Kg live weight,
normally slaughtered between November 2017 and January 2018 in abattoirs in the province of Teramo
(Italy) were taken into consideration.
\myparagraph{Scoring Pleurisy}
The scoring was performed by three veterinary surgeons, after a short training period during which they
agreed on how to interpret the different lesion classes. For each pig, the presence/absence and the
“severity” of pleurisy was recorded as follows:
\begin{itemize}
    \item A veterinarian was stationed where the post mortem inspection of organs is normally
performed. Inflammatory reactions of the visceral pleura (the pleural sheet lining the
pulmonary parenchyma) were recorded and quantified according to the SPES grid~\cite{dottori2007proposta};
    \item A second veterinarian was stationed at a later point of the slaughter chain and inspected the
parietal pleura (the pleural sheet which lines the chest wall).
\end{itemize}

Cases of pleurisy were quantified as follows: The parietal pleura was divided into three easily
identifiable areas (a) from the 1st to the 3rd intercostal spaces; (b) from the 4th to the 6th intercostal
spaces; (c) the remaining intercostal spaces, caudal to this point. In each of these areas, the presence
of pleural lesions was recorded, regardless of their extension, with the aim of limiting the
subjectivity of the score as much as possible. Considering the SPES system, and the fact that
lesions caused by Actinobacillus pleuropneumoniae usually involve both diaphragmatic lobes, the
following point system was given:
\begin{itemize}
    \item pleurisy effecting the first three intercostal spaces = 1 point;
    \item pleurisy effecting the 4th through 6th intercostal spaces = 2 points;
    \item pleurisy effecting the caudal intercostal spaces = 3 points.
\end{itemize}

Each pig was thus given a “total” score ranging between 0 and 12 which was obtained by adding the
individual scores of all the areas of the 2 half carcasses. This scoring method will be known herein with
the acronym PEPP (“Pleurisy Evaluation on Parietal Pleura”).

\myparagraph{Statistical Analysis}
The numerosity of the sample was determined for a generalised linear model, calculated with G*Power~\cite{faul2007g}. The averages of scores obtained with the two methods was compared based on
outcome (negative/positive) through one-way variance analysis. The relationship between the scores
obtained with the two methods was evaluated through a linear Pearson’s correlation coefficient (r). The
functional relationship between the variables measured with the two systems was resolved through
linear regression analysis, whose statistical significance was measured through variance analysis.

\section{Results}
\label{sec:results}
\begin{figure}[h]
\centering
\begin{tabular}{cc}
    \raisebox{-.15\height}{\includegraphics[width=0.35\columnwidth]{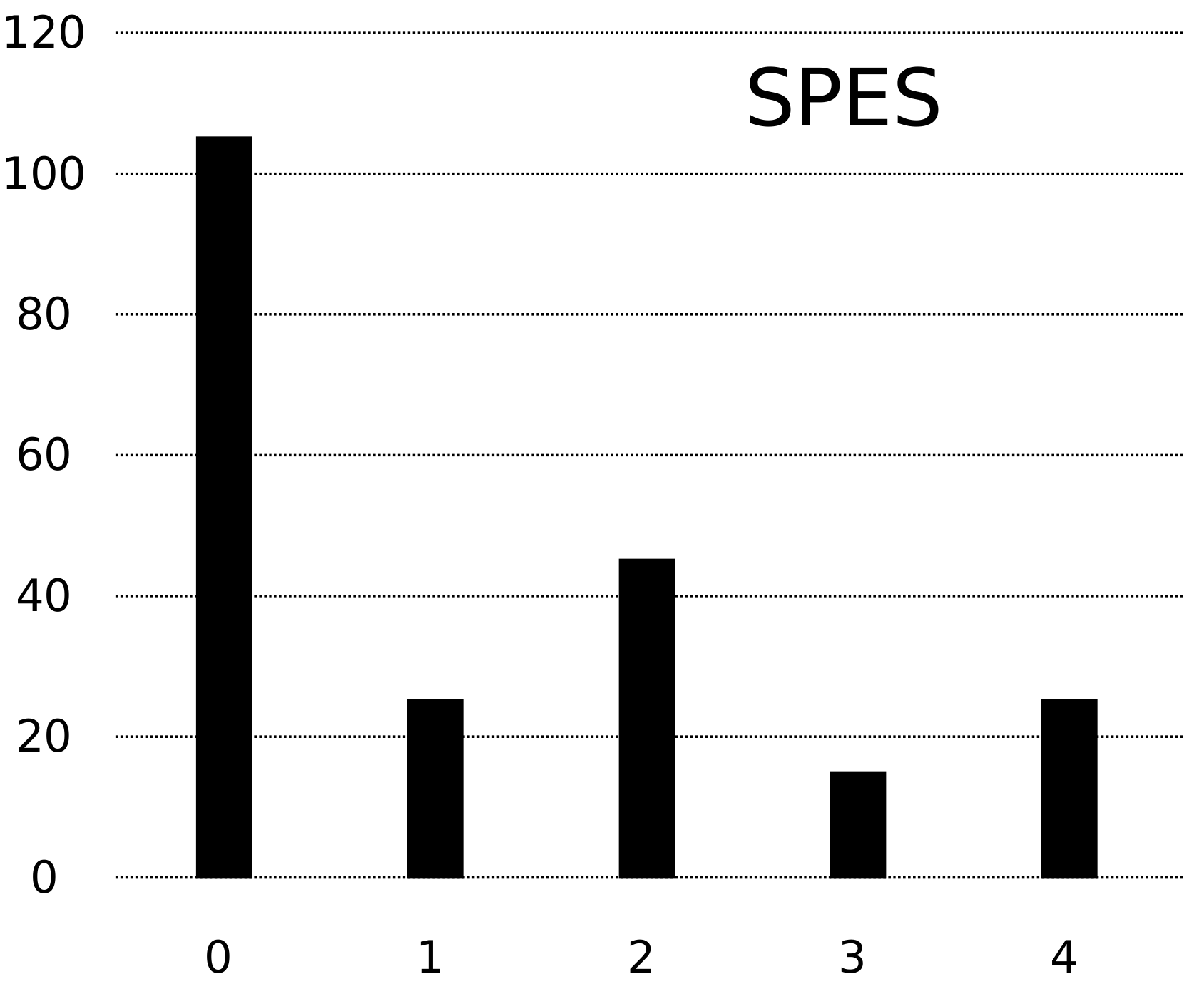}}
    &
    \hspace{0.3cm}
    \raisebox{-.15\height}{\includegraphics[width=0.6\columnwidth]{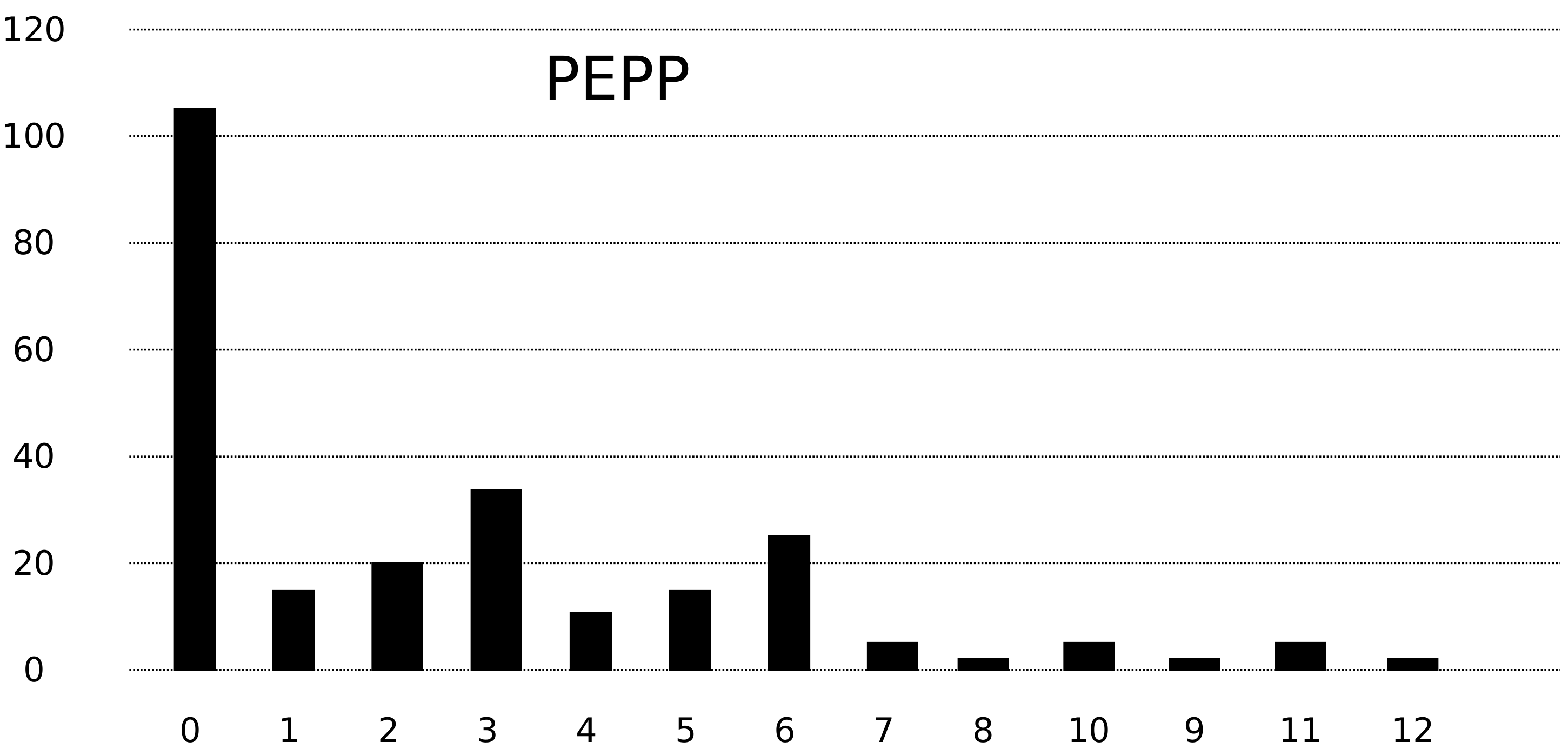}}
\end{tabular}
\caption{Graphical representation of the scores obtained by applying two different
pleurisy evaluation systems. Approximately, the 50\% of the pigs under study showed no
pleural lesion and obtained score 0 using both scoring systems. The distribution of scores
obtained by means of the SPES system was rather uniform, score 2 being more frequently
recorded. By applying the PEPP method, most of the pigs with pleural lesions fell in the score
interval 1-6, with a “tail” of pigs which scored >7.}
\label{fig:spes}
\end{figure}
\begin{figure}[t]
    \centering
    \includegraphics[width=0.8\textwidth]{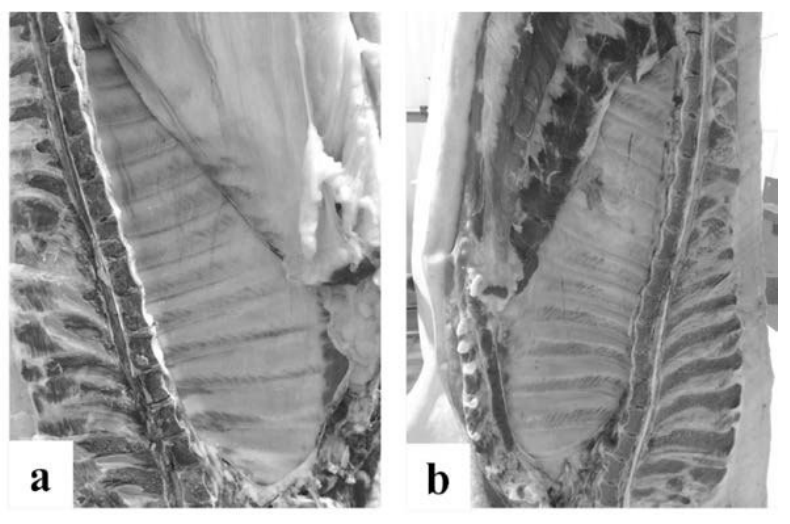}
    \caption{Inspection of the parietal pleura. The pictures show the healthy appearance of
the parietal pleura (a) and the presence of a large inlammatory reaction, easily recognizable
at the level of the caudal last intercostal spaces (b, score 3).}
\label{fig:spes_1}
\end{figure}

\begin{figure}[t]
    \centering
    \includegraphics[width=0.8\textwidth]{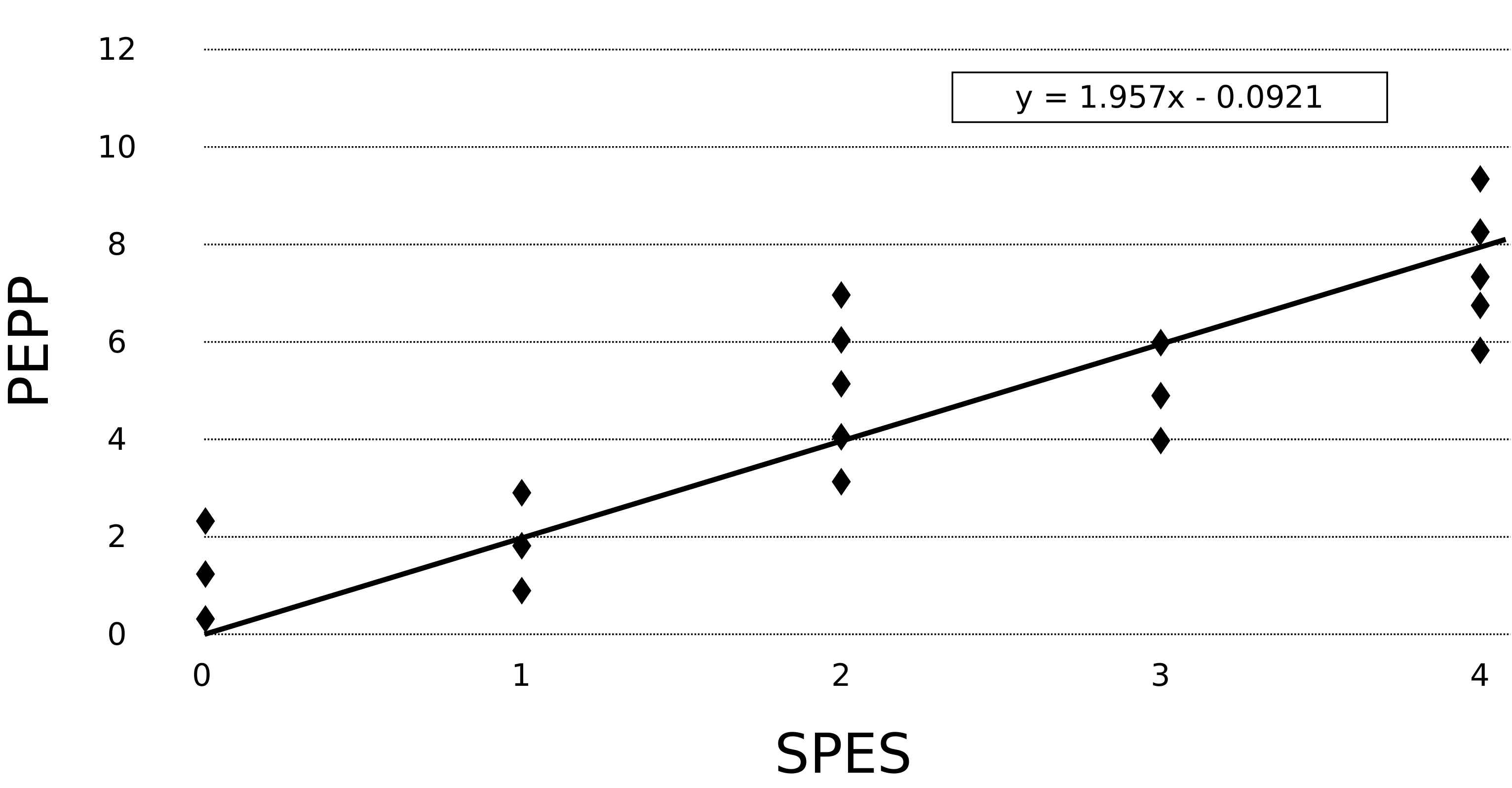}
    \caption{Analysis of the linear regression and equation between the two scoring systems
SPES and PEPP.}
\label{fig:spes_2}
\end{figure}

\myparagraph{Pleurisy scores using the SPES grid}
After the inspection of the lung and the visceral pleura, the presence of pleurisy was demonstrated in
109/216 pig (50,46\%), whereas no pleural inflammation was found in the remaining 107/216 pigs
(49,53\%). The distribution of scores obtained with the SPES method is shown in Figure 1.
\myparagraph{Pleurisy scores using the PEPP method}
The inspection of the parietal pleura and its scoring using the PEPP method required very little time (5-
10 sec/pig). The presence of inflammatory reactions on the parietal pleura was observed in 108/216
pigs (50\%), whereas the remaining 108/216 pigs (50\%) were considered “healthy”. Fig.~\ref{fig:spes_1} shows two examples. The
distribution of scores obtained with the PEPP method is shown in Fig.~\ref{fig:spes}.
\myparagraph{Statistical analysis}
The PEPP method proved capable of effectively differentiating “diseased” pigs from healthy ones
(p<0.01). The analysis of the correlation between results obtained with the two scoring methods show a
very high correlation coefficient (r = 0.913) and are statistically significant (p<0.01). Linear regression
analysis indicates that the determination coefficient is very high (R2 = 0.833) and statistically
significant (p<0.0001). Fig.~\ref{fig:spes_2} shows a graph with an analysis of the linear regression and equation
between the two scoring systems.
\subsection{Discussion}
Considering the usefulness of the slaughterhouse for monitoring the management of swine farms, and
more generally modern animal production, new evaluation methods, which more fully adapt to the
specific needs of each production system are a positive development. In this sense, the PEPP method,
shows itself as a possible alternative to the SPES system, which has been amply documented and
recognized on a worldwide scale. As expected, the two scoring methods gave overlapping results, with
an extremely high correlation coefficient. In fact, with rare exceptions (for example limited extension
of pleurisy, circumscribed to the interlobar spaces), such lesions normally involve both pleural sheets
(visceral and parietal). A confirmation of this was seen in the almost complete overlapping of pigs
considered “healthy” (score 0) using both evaluation methods. In fact, only eight pigs with interlobar
adherences were not correctly identified as “diseased” through the application of the PEPP method;
whereas seven pigs with small lesions confined to the first few intercostal spaces were not correctly
identified as “diseased” through the application of the SPES grid. Obviously the PEPP method has both
pros and cons. for example, the inspection of the carcass halves at the end of the slaughterchain does
not permit the veterinarian to inspect other lesions (es. Pneumonia, pericarditis, hepatitis). On the other
hand, this system permits the veterinarian to perform the scoring in a cleaner and more relaxed
environment, in a potentially extremely short period of time. In addition, the scoring of the parietal
pleura is simpler, and less influenced by other factors (es. Blood staining, pulmonary scarring, etc.) and
can potentially be performed at a later time through the analysis of digital images. To this aim other
studies are currently being carried out, to try to standardise the scoring between veterinarians, using
different systems both in the slaughterhouse, and remotely on digital images.
\section{Conclusions}
\label{sec:conclusions}
The inspection of the parietal pleura and the scoring of any lesions present proved simple and rapidly
applicable. It is herein proposed as an alternative to the SPES system.
\myparagraph{Acknowledgement}
We kindly thank Mr. Andrea Paolini and Mr. Vittorio Mosca, students at the faculty of Veterinary
Medicine of Teramo for their help with taking pictures at the slaughterhouse

\bibliographystyle{splncs04}

\end{document}